\documentclass[conference]{IEEEtran}
\usepackage{cite}
\usepackage{amsmath,amssymb,amsfonts}
\usepackage{algorithmic}
\usepackage{graphicx}
\usepackage{textcomp}
\usepackage{xcolor}
\usepackage{amsmath}
\usepackage{amsfonts}
\usepackage{url}
\def\BibTeX{{\rm B\kern-.05em{\sc i\kern-.025em b}\kern-.08em
    T\kern-.1667em\lower.7ex\hbox{E}\kern-.125emX}}
\begin{document}

\title{DeepCrysTet: A Deep Learning Approach\\Using Tetrahedral Mesh for Predicting\\Properties of Crystalline Materials}

\author{\IEEEauthorblockN{Hirofumi Tsuruta$^1$, Yukari Katsura$^{2,3,4}$, Masaya Kumagai$^{1,4,5}$}
\IEEEauthorblockA{
$^1$\textit{SAKURA internet Inc.}, 
$^2$\textit{National Institute for Materials Science (NIMS)}, \\
$^3$\textit{The University of Tokyo}, 
$^4$\textit{RIKEN}, 
$^5$\textit{Kyoto University} \\
hi-tsuruta@sakura.ad.jp}
}

\maketitle

\begin{abstract}
Machine learning (ML) is becoming increasingly popular for predicting material properties to accelerate materials discovery.
Because material properties are strongly affected by its crystal structure, a key issue is converting the crystal structure into the features for input to the ML model.
Currently, the most common method is to convert the crystal structure into a graph and predicting its properties using a graph neural network (GNN).
Some GNN models, such as crystal graph convolutional neural network (CGCNN) and atomistic line graph neural network (ALIGNN), have achieved highly accurate predictions of material properties.
Despite these successes, using a graph to represent a crystal structure has the notable limitation of losing the crystal structure's three-dimensional (3D) information.
In this work, we propose DeepCrysTet, a novel deep learning approach for predicting material properties, which uses crystal structures represented as a 3D tetrahedral mesh generated by Delaunay tetrahedralization.  
DeepCrysTet provides a useful framework that includes a 3D mesh generation method, mesh-based feature design, and neural network design.
The experimental results using the Materials Project dataset show that DeepCrysTet significantly outperforms existing GNN models in classifying crystal structures and achieves state-of-the-art performance in predicting elastic properties.
\end{abstract}

\begin{IEEEkeywords}
Deep Learning, Neural Network, Materials Informatics, Crystal Structure, Tetrahedral Mesh
\end{IEEEkeywords}

\section{Introduction}

Machine learning (ML) has emerged as a powerful tool for accelerating the design of new materials~\cite{choudhary_2022}.
The primary application of ML is to predict material properties based on chemical and structural information without time-consuming experiments.
Conventional methods for predicting material properties involve the application of ab initio calculations by applying, amongst others, density functional theory (DFT), which has proven computationally expensive.
By the adoption of ML, predictions are expected to be significantly faster than DFT calculations, leading to the efficient design of new materials with desired properties.

\begin{figure}[tb]
  \centering
  \includegraphics[width=\linewidth]{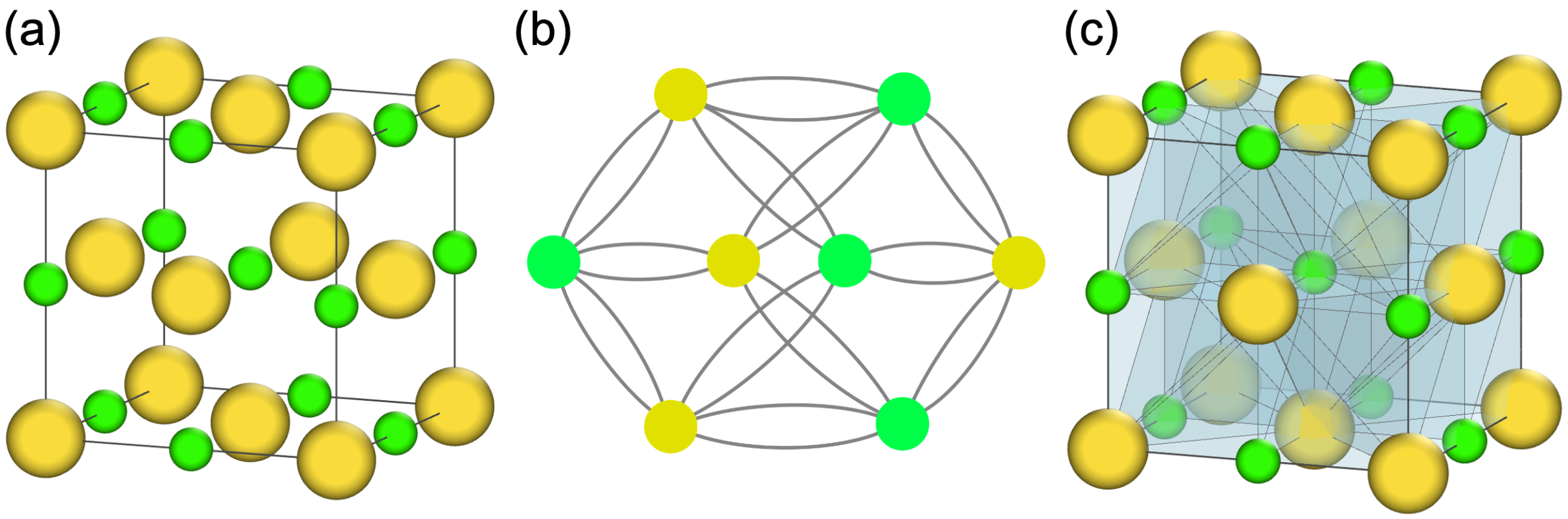}
  \caption{Illustration of a crystal structure of NaCl with three representations: (a) a 3D point cloud, (b) a graph used in existing studies~\cite{xie_2018,chen_2019,park_2020,choudhary_2021}, and (c) a 3D tetrahedral mesh used in our work. The cubes in (a) and (c) illustrate unit cells.} 
  \label{fig:crystal_structure}
\end{figure}

Because most inorganic materials, such as metals and ceramics, are crystalline and their properties are strongly affected by their crystal structures, a key issue in predicting the properties of crystalline materials is the conversion of the crystal structure into features for input to the ML model.
Currently, the most common and practical method is to convert the crystal structure into a graph, as illustrated in Fig.~\ref{fig:crystal_structure} (b).
To date, various graph neural network (GNN) architectures have been rapidly developed to predict the properties of crystalline materials~\cite{xie_2018,chen_2019,park_2020,choudhary_2021}.
The crystal graph convolutional neural network (CGCNN)~\cite{xie_2018}, a representative model, has achieved highly accurate predictions of properties such as formation energy and band gap.

Despite these successes, using a graph to represent a crystal structure has the notable limitation of losing the crystal structure's three-dimensional (3D) information.
The crystal structure is originally composed of 3D atomic coordinates and a space lattice representing the 3D periodicity of these coordinates, as shown in Fig.~\ref{fig:crystal_structure} (a).
It is known that the 3D atomic coordinates and the 3D shape of the unit cell, the periodic unit of the space lattice, are related to certain material properties, such as magnetic~\cite{rungger2006ab} and elastic~\cite{furmanchuk2016predictive} properties.
Therefore, prediction accuracy can be further improved by designing a data representation that takes advantage of the 3D crystal structure.

In this work, we propose DeepCrysTet, a deep learning-based approach for predicting material properties, which uses crystal structures represented as a 3D tetrahedral mesh to overcome the limitations of graph-based approaches.
DeepCrysTet provides a valuable framework comprising three core components: a 3D mesh generation method, mesh-based feature design, and neural network design.
The 3D tetrahedral mesh of the crystal structure used in DeepCrysTet, as shown in Fig.~\ref{fig:crystal_structure} (c), is generated by Delaunay tetrahedralization~\cite{delaunay_1934}.
Furthermore, we designed two types of features: structural features of the crystal structure and chemical features of constituent atoms, using triangular faces in the tetrahedral mesh as a basic unit.
Finally, we designed a neural network that predicts various material properties using a set of triangular faces as inputs.

The main contributions of this paper are summarized as follows.

\begin{itemize}
  \setlength{\itemsep}{2pt}
  \item To our knowledge, DeepCrysTet is the first work to predict material properties using a 3D mesh representation of a crystal structure.
  \item The framework of DeepCrysTet, including the 3D mesh generation method, mesh-based feature design, and neural network design, can be utilized as a promising alternative to the mainstream approach that uses graphs for predicting material properties.
  \item The experimental results show that DeepCrysTet significantly outperforms existing GNN approaches in classifying crystal structures and achieves state-of-the-art performance in predicting elastic properties.
\end{itemize}

\section{Related Work}

\subsection{Prediction of Crystalline Material Properties}

ML has been increasingly employed to predict the properties of crystalline materials~\cite{choudhary_2022}.
A critical issue in the application of ML is the conversion of the crystal structure into features that are compatible with the ML model.
There are currently two main approaches to converting crystal structures: hand-crafted descriptors~\cite{ward_2016,isayev_2017,choudhary_2018} and graph representation~\cite{xie_2018,chen_2019,park_2020,choudhary_2021}.
Although ML models based on hand-crafted descriptors have achieved some success in predicting properties, these descriptors are designed by humans with specific domain knowledge, which inherently limits them.

In recent years, GNNs have attracted considerable attention for predicting the properties of crystalline materials, offering more substantial performance improvements than approaches based on hand-crafted descriptors.
The CGCNN~\cite{xie_2018} is a pioneering GNN architecture for accurately predicting the properties of crystalline materials.
CGCNN builds a convolutional neural network directly on a graph with one node for each constituent atom and edges corresponding to the originally defined interatomic connections.
Recent variants, such as materials graph network (MEGNet)~\cite{chen_2019}, improved crystal graph convolutional neural network (iCGCNN)~\cite{park_2020}, and atomistic line graph neural network (ALIGNN)~\cite{choudhary_2021}, have been developed to improve predictive performance.
MEGNet improved CGCNN by incorporating global state inputs, including temperature, pressure, and entropy.
In another effort, iCGCNN outperformed CGCNN by incorporating information from the Voronoi tessellated crystal structure.
Notably, ALIGNN, which explicitly encodes the bond angle information, is currently among the state-of-the-art models with the best performance on the Materials Project dataset~\cite{jain_2013}.

\subsection{Deep Learning on 3D Mesh}

A 3D mesh consists of a collection of vertices, edges, and faces, and is currently a widely used data representation in fields such as computer vision and computer graphics.
Vertices are connected by edges that define connectivity, indicating how each vertex is connected to the other.
The resulting closed sets of edges form the faces.
Owing to the combination of the geometric structures of vertices, edges, and faces, a 3D mesh can be more expressive than other 3D representations, such as point clouds and voxels.

However, dealing with 3D meshes is difficult because they consist of multiple elements, and the permutations of these elements are arbitrary.
Recently, considerable effort has been devoted to developing deep learning models for 3D mesh data to address these difficulties.
MeshNet~\cite{feng_2019} is a pioneering model on 3D mesh that treats the mesh face as a basic unit and extracts the spatial and structural features of each face.
MeshNet adopts per-face processes and symmetric functions, which are similar to PointNet~\cite{qi2017pointnet}, to satisfy the permutation invariant of elements in mesh data.
MeshNet++~\cite{singh_2021} is a deep learning model designed to learn local structures at multiple scales and achieves state-of-the-art performance over MeshNet in 3D shape classification.

\section{DeepCrysTet: Deep Learning Approach Using Tetrahedral Mesh}

\begin{figure*}[tb]
  \centering
  \includegraphics[width=0.85\textwidth]{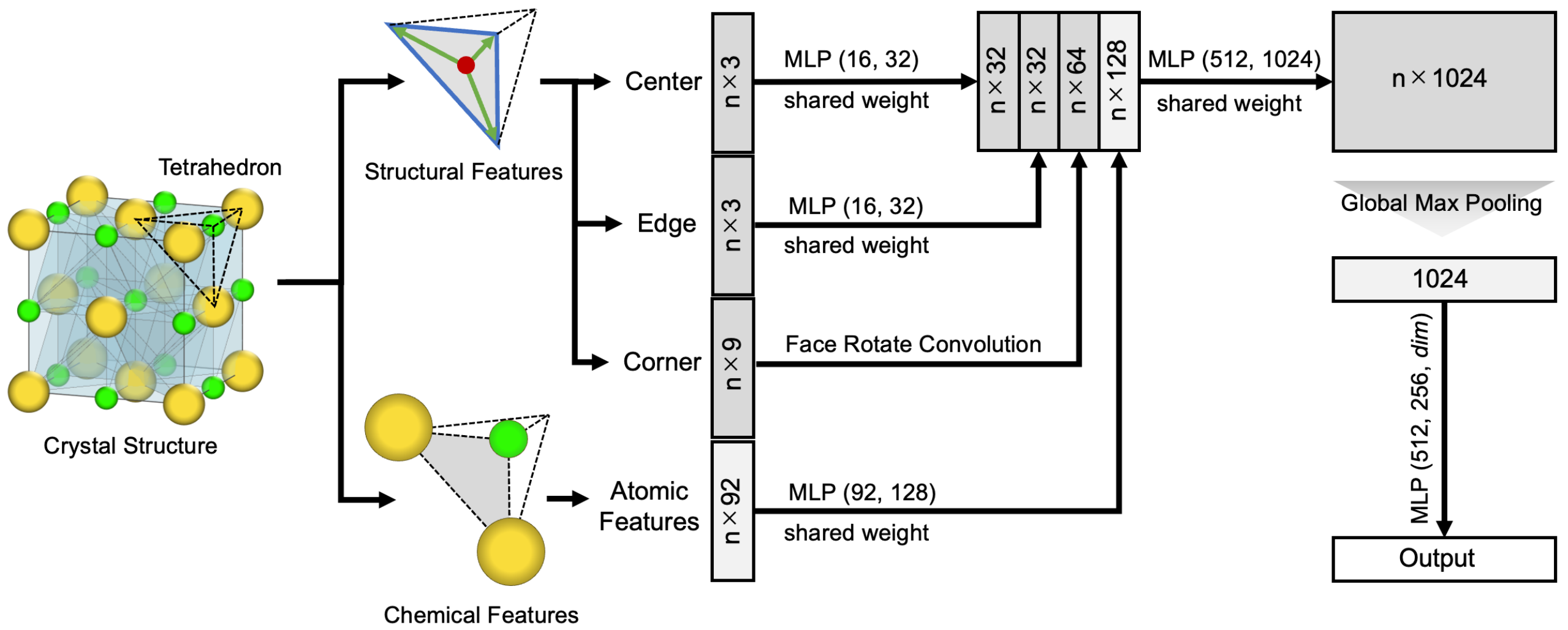}
  \caption{Overall framework of DeepCrysTet.} 
  \label{fig:framework}
\end{figure*}

In this section, we present DeepCrysTet, a deep learning approach that uses a crystal structure represented as a 3D tetrahedral mesh for predicting material properties.
The overall framework of DeepCrysTet is shown in Fig.~\ref{fig:framework}.
First, we introduce a method for converting a crystal structure into a 3D mesh.
Then, we describe the design of the input features derived from the generated mesh.
Finally, we explain the proposed neural network architecture for addressing the designed features.

\subsection{Mesh Generation}

We adopted Delaunay tetrahedralization~\cite{delaunay_1934} to generate a 3D tetrahedral mesh of the crystal structure.
Delaunay tetrahedralization divides a 3D space into a set of tetrahedra from a given set of 3D vertices.
The resulting tetrahedral mesh was filled with tetrahedra without gaps inside the crystal structure, as shown in Fig.~\ref{fig:crystal_structure} (c).
Although Delaunay tetrahedralization is a fundamental technique in various applications, such as finite element analysis and computer graphics, few studies have applied it to crystal structures~\cite{hinuma_2022}.

The crystal structure originally comprises 3D atomic coordinates and a space lattice representing the 3D periodicity of these coordinates.
A periodic unit in the space lattice is termed a unit cell and is represented by a cube as in Fig.~\ref{fig:crystal_structure} (a).
The periodicity of the crystal structure enables us to express the structural features of the entire crystal structure by a collection of tetrahedra that exist near the unit cell.

Here, we describe concrete methods for generating the mesh.
First, point cloud data were created from the 3D coordinates of each atom contained in 5$\times$5$\times$5 unit cells.
Next, Delaunay tetrahedralization was applied to the point cloud data, and the set of tetrahedra contained in one unit cell located at the center of 5$\times$5$\times$5 unit cells was extracted as the mesh data to be used.
If the central unit cell contained at least one of the four vertices of a tetrahedron, the tetrahedron was extracted. 
Note that we allowed a deviation of $\pm$0.05 \AA~ to determine whether a vertex is in the unit cell.
Finally, the vertices in the mesh of each crystal structure were parallel-shifted such that the centroid of the vertices coincided with the coordinate origin.

\subsection{Feature Design}
\label{sec:feature_design}

\begin{figure*}[tb]
  \centering
  \includegraphics[width=0.7\textwidth]{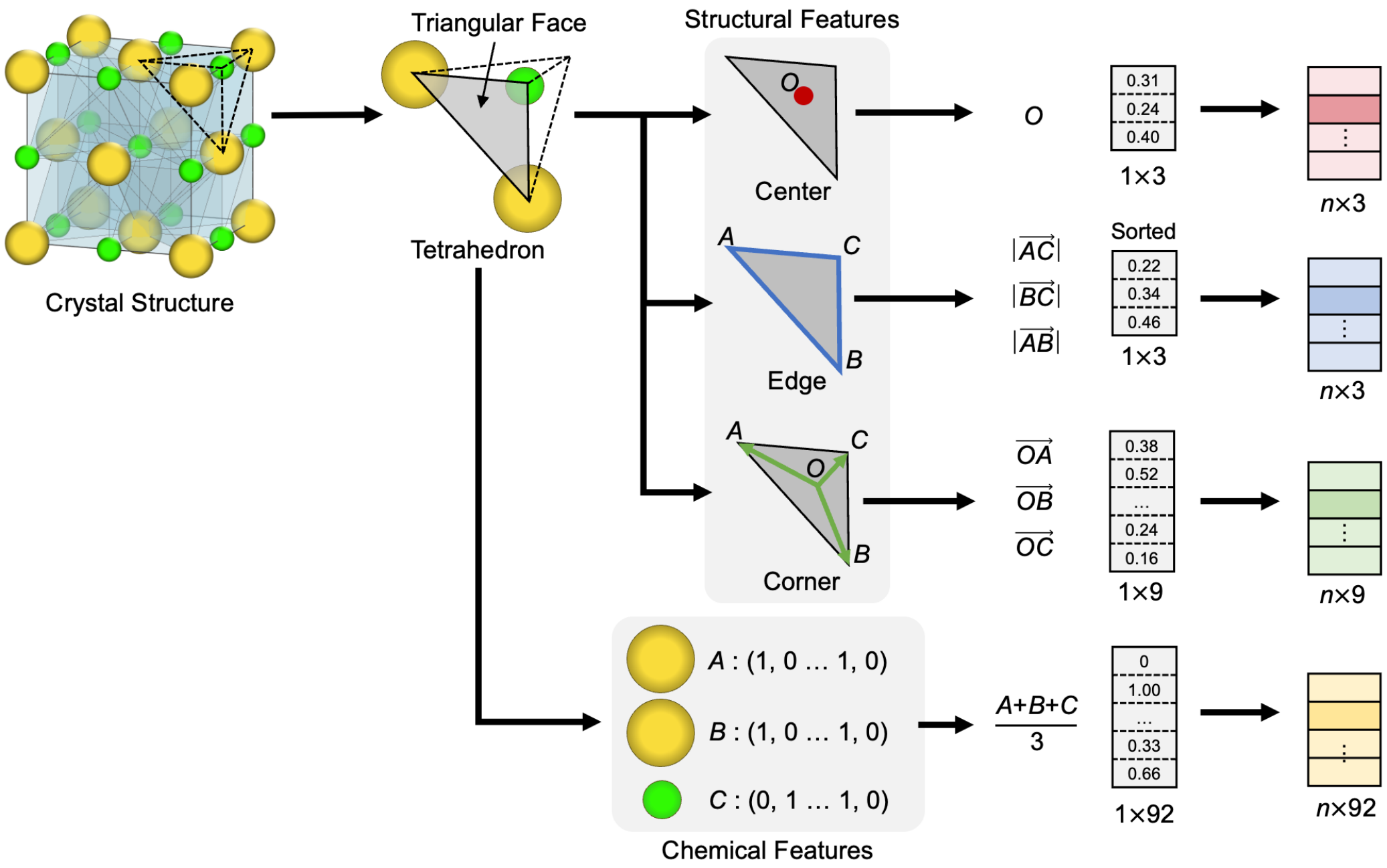}
  \caption{Design of the input features derived from a tetrahedral mesh.}
  \label{fig:mesh_features}
\end{figure*}

In this section, we elaborate on the design of input features derived from a tetrahedral mesh, as illustrated in Fig.~\ref{fig:mesh_features}.
The mesh data generated by Delaunay tetrahedralization is a set of tetrahedra.
Because each tetrahedron consists of four triangular faces, the mesh data can also be regarded as a set of triangular faces.
To take advantage of a mesh that contains 3D connectivity between constituent atoms, we designed features of two types, structural and chemical, using the triangular faces as the basic units.
For each face, we defined the following four features.

\begin{itemize}
  \setlength{\itemsep}{2pt}
  \item Structural features
  \begin{itemize}
    \item Center: Coordinates of the centroid of the triangle.
    \item Edge: Lengths of triangle sides sorted in ascending order.
    \item Corner: Vectors from the centroid to three vertices.
  \end{itemize}
  \item Chemical features
  \begin{itemize}
    \item Atomic features: Average of the feature vectors of the three atoms constituting the triangle.
  \end{itemize}
\end{itemize}

As features for the atoms, we used a 92-dimensional binary vector discretized from the nine atomic features used in the CGCNN, as shown in Table~\ref{tab:atom_features}.

\begin{table}[tb]
  \caption{Properties used in atomic features.}
  \centering
  \renewcommand{\arraystretch}{1.1}
  \setlength{\tabcolsep}{3pt}
  \begin{tabular}{cccc} \hline
    Property & Unit & Range & \#categories \\ \hline\hline
    Group number & - & 1, 2, ..., 18 & 18 \\
    Period number & - & 1, 2, ..., 9 & 9 \\
    Electronegativity & - & 0.5--4.0 & 10 \\
    Covalent radius & pm & 25--250 & 10 \\
    Valence electrons & - & 1, 2, ..., 12 & 12 \\
    First ionization energy & eV & 1.3--3.3 & 10 \\
    Electron affinity & eV & $-$3--3.7 & 10 \\
    Block & - & s, p, d, f & 4 \\
    Atomic volume & cm$^3$/mol & 1.5--4.3 & 10 \\ \hline
 \end{tabular}
  \label{tab:atom_features}
\end{table}

\subsection{Neural Network Architecture}

In this section, we describe the neural network architecture of the model shown in Fig.~\ref{fig:framework}.
As the model's input is set data, the model should satisfy two critical requirements~\cite{zaheer2017deep,lee2019set}.
First, it must be permutation invariant.
Otherwise expressed, even if the permutation of the input set elements changes, the model's output must not change.
Second, the model should process input sets of different sizes.
The mesh data used in this work have different numbers of triangular faces for each material.

To satisfy these two requirements, the proposed model is based on the PointNet architecture~\cite{qi2017pointnet}.
Given an unordered set $\{x_1, x_2, ..., x_n\}$ with $x_i \in \mathbb{R}^d$, we can define a model function $f$ that maps a set to a vector as

\begin{equation}
  \label{eq:symmetry_function}
  f(\{x_1, x_2, ..., x_n\}) = g\Bigl(\underset{i=1,...,n}{MAX}\{h(x_i)\}\Bigr)
\end{equation}

where $g$ and $h$ are the multi-layer perceptron (MLP) and $MAX$ is a max pooling operation over the elements of the set.
First, each element in a set is independently processed by the same function $h$, that is, the weights of $h$ are shared for each element.
The max pooling operation then processes the outputs of $h$ to aggregate the information from all elements.
Note that the output obtained through the max pooling operation is a fixed-dimensional vector regardless of the set size $n$.
The final output is obtained using the MLP $g$.
The model function $f$ satisfies both requirements, as mentioned above.

Next, we present the detailed design of DeepCrysTet based on Equation (\ref{eq:symmetry_function}).
As described in section \ref{sec:feature_design}, we designed four features from each set element, that is, a triangular face, and each of the four features is initially processed independently.
The center, edge, and atomic features are processed by the MLP with shared weights for each element.
The corner is processed by the face rotate convolution proposed in MeshNet~\cite{feng_2019}, which captures the inner structure of the faces.
Each of the four processed features is combined and processed by the MLP with shared weights and is then transformed into a vector with fixed dimensions by the max pooling operation.
Finally, the output value is obtained by processing with the MLP.

\section{Experiments}

We experimentally applied DeepCrysTet to two tasks: crystal structure classification and material property prediction.
First, we evaluated the classification accuracies of the crystal systems and space groups to confirm whether DeepCrysTet can capture 3D crystal structures.
Next, we evaluated the prediction errors of the four material properties.

\subsection{Experimental Settings}

\subsubsection{Dataset.}

We used the 2018.10.18 version\footnote{Materials Project Data: \url{https://figshare.com/articles/dataset/Materials_Project_Data/7227749}} of the Materials Project dataset~\cite{jain_2013}.
The dataset contained the crystal structures and various types of properties of 83,989 crystalline materials.
We selected six properties: the crystal system, space group, formation energy, band gap, bulk modulus, and shear modulus.
Regarding the bulk and shear moduli, missing and negative values were removed, respectively leaving 7,643 and 7,432 remaining materials.

\subsubsection{Baselines.}

We compared the performance of DeepCrysTet with the following two baselines:

\begin{itemize}
  \setlength{\itemsep}{2pt}
  \item CGCNN~\cite{xie_2018}: CGCNN is a pioneering GNN model for predicting material properties using crystal graphs.
  \item ALIGNN~\cite{choudhary_2021}: ALIGNN is the latest GNN model inspired by CGCNN and currently achieves state-of-the-art performance on the Materials Project dataset by incorporating bond angle information.
\end{itemize}

We adopted the hyperparameters reported in the original papers for both baselines.
We used the original implementations of CGCNN\footnote{CGCNN: \url{https://github.com/txie-93/cgcnn}} and ALIGNN\footnote{ALIGNN: \url{https://github.com/usnistgov/alignn}} provided by the authors.

\subsubsection{Model Training.}

The models were trained for 200 epochs using the Adam optimizer and an initial learning rate of 0.001.
The batch size was 32 for the bulk and shear moduli and 128 for all other properties.
We randomly split the dataset into training, validation, and test sets at a ratio of 80:10:10.
All the results presented here are based on the performance of the 10 \% of the test sets that were not used during the training or model selection procedures.

\subsubsection{Implementation.}
We used the Qhull library~\cite{barber_1996} provided by Scipy~\cite{virtanen_2020} for Delaunay tetrahedralization and the Pymatgen library~\cite{ong_2013} to handle and generate crystal structures, respectively.
We implemented our neural network model using PyTorch~\cite{paszke2019pytorch}.
Our source code is publicly available at \url{https://github.com/tsurubee/DeepCrysTet}.

\subsection{Classification of Crystal Structure}

\begin{table}[tb]
  \caption{Summary of the classification accuracies for crystal systems and space groups on the test sets. The best performance is highlighted in bold.}
  \centering
  \renewcommand{\arraystretch}{1.2}
  \setlength{\tabcolsep}{4pt}
  \begin{tabular}{ccccc} \hline
    Classification & \#classes & DeepCrysTet & CGCNN & ALIGNN \\ \hline\hline
    Crystal systems & 7 & \bf{97.5 \%} & 63.4 \% & 75.6 \% \\
    Space groups & 230 & \bf{90.3 \%} & 49.9 \% & 64.3 \% \\ \hline
  \end{tabular}
  \label{tab:results_classification}
\end{table}

\begin{figure*}[tb]
  \centering
  \includegraphics[width=0.75\textwidth]{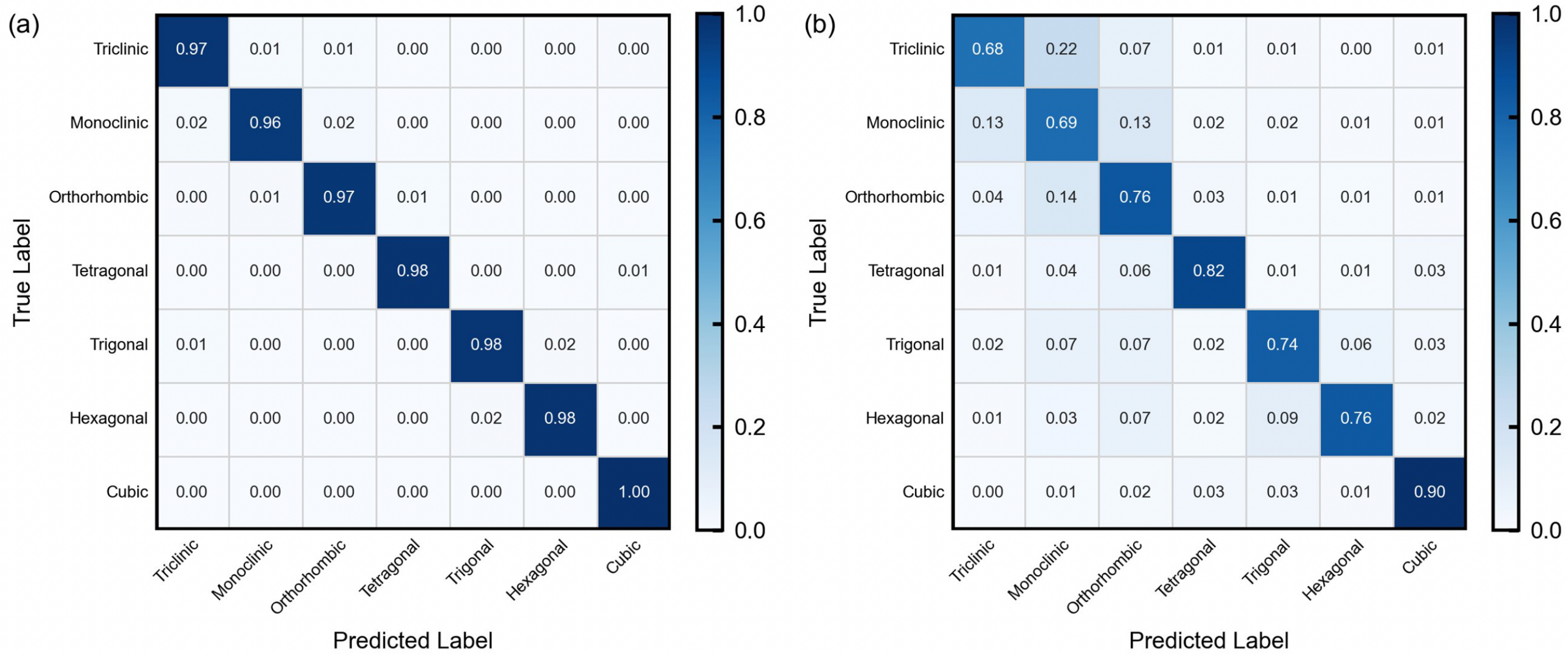}
  \caption{Confusion matrices of (a) DeepCrysTet and (b) ALIGNN for the classification of crystal systems.} 
  \label{fig:confusion_matrices}
\end{figure*}

We evaluated the classification accuracies of the crystal systems and space groups to confirm whether DeepCrysTet can capture 3D crystal structures.
The crystal structures are classified into seven crystal systems based on the 3D shape of the unit cell.
The crystal structures are further classified into 230 space groups based on the symmetry of the 3D atomic coordinates in the unit cell.
These two classifications strongly reflect the 3D crystal structure.

Table~\ref{tab:results_classification} lists the classification accuracies of DeepCrysTet and the two baselines.
The classification accuracies of DeepCrysTet for the crystal systems and space groups were 97.5 \% and 90.3 \%, respectively, which significantly exceeded those of CGCNN and ALIGNN.
This result indicates that the graph representation used in the baseline models has difficulty capturing the 3D crystal structure.
Although the better accuracy of ALIGNN compared to CGCNN implies that the bond angle information has a positive effect on capturing the 3D crystal structure, its accuracy did not match DeepCrysTet.

Fig.~\ref{fig:confusion_matrices} shows the confusion matrices normalized over the true labels of DeepCrysTet and ALIGNN for the classification of the crystal systems.
As shown in Fig.~\ref{fig:confusion_matrices} (a), DeepCrysTet achieved almost equally high classification accuracies for all classes.
In contrast, ALIGNN exhibits variations in the classification accuracy for each class, as shown in Fig.~\ref{fig:confusion_matrices} (b).
Furthermore, we found that ALIGNN is likely to misclassify the triclinic and monoclinic systems.
Several previous studies~\cite{suzuki2020symmetry,li2021composition} have also shown that it is more difficult to classify the triclinic and monoclinic systems than others, which is consistent with ALIGNN's tendency.
Conversely, DeepCrysTet succeeded in classifying the triclinic and monoclinic systems with high accuracy.
These results demonstrate that DeepCrysTet can capture 3D crystal structures more accurately from 3D mesh representations than graph-based approaches.

\subsection{Prediction of Material Properties}

\begin{table}[tb]
  \caption{Summary of the prediction performance (MAE) for four different properties on the test sets. The best performance is highlighted in bold.}
  \centering
  \renewcommand{\arraystretch}{1.2}
  \setlength{\tabcolsep}{4pt}
  \begin{tabular}{ccccc} \hline
    Property & Unit & DeepCrysTet & CGCNN & ALIGNN \\ \hline\hline
    Formation energy & eV/atom & 0.062 & 0.043 & \bf{0.024} \\
    Band gap & eV & 0.279 & 0.271 & \bf{0.211} \\
    Bulk modulus & log(GPa) & \bf{0.061} & 0.076 & 0.068 \\
    Shear modulus & log(GPa) & \bf{0.106} & 0.129 & 0.115 \\ \hline
  \end{tabular}
  \label{tab:results_regression}
\end{table}

We evaluated the prediction performance of DeepCrysTet and two baselines using four material properties as target variables: formation energy, band gap, bulk modulus, and shear modulus.
In this experiment, the mean absolute error (MAE) was used as the evaluation metric.
Table~\ref{tab:results_regression} lists the experimental results of the prediction performance of DeepCrysTet and the two baselines.
The two baselines are shown to outperform DeepCrysTet in predicting the formation energy and band gap.
However, the predictive performance of DeepCrysTet for the two elastic properties, the bulk and shear moduli, exceeds that of the two baselines.
These results are probably due to the data representation of the crystal structure between DeepCrysTet and the baselines, being entirely different.
In addition, the experimental results suggest that the 3D crystal structure captured by DeepCrysTet contributes more to predicting the elastic properties than the other properties.

Here, we discuss why DeepCrysTet is superior in predicting elastic properties.
Previous studies~\cite{rabiei2020measurement,rabiei2021relationship} have shown that the elastic modulus is strongly correlated with the planar density, which is the fraction of the crystal plane area occupied by atoms.
The planar density is determined by the plane area and radii of the constituent atoms.
Although the crystal plane is not the same as the triangular face used in DeepCrysTet, both are two-dimensional planes formed by adjacent atoms in a 3D crystal structure.
Furthermore, the features used in DeepCrysTet include the size and shape of each triangular face and radii of the constituent atoms, which can be regarded as features similar to the planar density.
Hence, these features may contribute to the predictive performance of elastic properties.

\section{Conclusion}

We propose DeepCrysTet, a novel deep learning approach for predicting the properties of crystalline materials.
To address the limitations of existing graph-based approaches, the proposed method focuses on representing the crystal structure as a 3D tetrahedral mesh acquired through Delaunay tetrahedralization.
To our knowledge, DeepCrysTet is the first work to predict material properties using a 3D mesh representation of a crystal structure.
To evaluate the model's performance, two tasks: crystal structure classification and material property prediction, were performed using the Materials Project dataset.
DeepCrysTet enables accurate classification into crystal systems and space groups, which is difficult to achieve with current GNN models thereby demonstrating that DeepCrysTet can capture 3D crystal structures.
Furthermore, DeepCrysTet achieved a prediction accuracy comparable to that of state-of-the-art GNN models for predicting the elastic properties of materials.
In the future, we plan to improve DeepCrysTet further to achieve highly accurate predictions of the formation energy and band gap in addition to the elastic properties.
Because current GNN models have demonstrated the importance of convolutional layers in aggregating neighboring node features, we will explore the design of convolutional layers to aggregate features from neighboring triangular faces in a 3D tetrahedral mesh.

\section*{Acknowledgment}

This work was supported by JST CREST Grant Number JPMJCR19J1 and JSPS KAKENHI Grant-in-Aid for Early-Career Scientists Grant Number JP22K14474.

\bibliographystyle{IEEEtran}

\end{document}